\documentclass[acmsmall,nonacm]{acmart}

\AtBeginDocument{%
  \providecommand\BibTeX{{%
    \normalfont B\kern-0.5em{\scshape i\kern-0.25em b}\kern-0.8em\TeX}}}

\setcopyright{acmcopyright}
\copyrightyear{2018}
\acmYear{2018}
\acmDOI{10.1145/1122445.1122456}

\acmConference[HATRA '20]{HATRA '20: Human Aspects of Types and Reasoning Assistants}{November 15--20, 2020}{Online Workshop}
\acmPrice{15.00}
\acmISBN{978-1-4503-XXXX-X/18/06}

\usepackage{booktabs}
\usepackage{tabularx}
\usepackage{listings}
\usepackage{cleveref}

\newcommand{\code}[1]{\texttt{#1}}

\begin{document}

\title{User-Centered Programming Language Design: A Course-Based Case Study}

\author{Michael Coblenz}
\orcid{0000-0002-9369-4069}             
\affiliation{
  \position{Postdoctoral Fellow}
  \department{Computer Science Department}              
  \institution{University of Maryland}            
  \streetaddress{8125 Paint Branch Drive}
  \city{College Park}
  \state{MD}
  \postcode{20742}
  \country{USA}                    
}
\email{mcoblenz@cs.cmu.edu}          

\author{Ariel Davis}
\affiliation{
    \department{Information Networking Institute}
    \institution{Carnegie Mellon University}            
    \streetaddress{5000 Forbes Ave.}
    \city{Pittsburgh}
    \state{PA}
    \postcode{15213}
    \country{USA} 
}
\email{azdavis@andrew.cmu.edu}

\author{Megan Hofmann}
\affiliation{
    \department{Human-Computer Interaction Institute}
    \institution{Carnegie Mellon University}            
    \streetaddress{5000 Forbes Ave.}
    \city{Pittsburgh}
    \state{PA}
    \postcode{15213}
    \country{USA} 
}
\email{hofmann.megan@gmail.com}

\author{Vivian Huang}
\affiliation{
    \institution{Carnegie Mellon University}            
    \streetaddress{5000 Forbes Ave.}
    \city{Pittsburgh}
    \state{PA}
    \postcode{15213}
    \country{USA} 
}
\email{vivianh@andrew.cmu.edu}

\author{Siyue Jin}
\affiliation{
    \department{Information Networking Institute}
    \institution{Carnegie Mellon University}            
    \streetaddress{5000 Forbes Ave.}
    \city{Pittsburgh}
    \state{PA}
    \postcode{15213}
    \country{USA} 
}
\email{siyuej@alumni.cmu.edu}

\author{Max Krieger}
\affiliation{
    \position{Undergraduate Student}
    \department{Psychology}
    \institution{Carnegie Mellon University}            
    \streetaddress{5000 Forbes Ave.}
    \city{Pittsburgh}
    \state{PA}
    \postcode{15213}
    \country{USA} 
}
\email{mfkriege@andrew.cmu.edu}

\author{Kyle Liang}
\affiliation{
    \position{Ph.D. Student}
    \department{Institute for Software Research}
    \institution{Carnegie Mellon University}            
    \streetaddress{5000 Forbes Ave.}
    \city{Pittsburgh}
    \state{PA}
    \postcode{15213}
    \country{USA} 
}
\email{kmliang@andrew.cmu.edu}

\author{Brian Wei}
\affiliation{
    \position{Undergraduate Student}
    \department{Computer Science}
    \institution{Carnegie Mellon University}            
    \streetaddress{5000 Forbes Ave.}
    \city{Pittsburgh}
    \state{PA}
    \postcode{15213}
    \country{USA} 
}
\email{bwei1@andrew.cmu.edu}

\author{Mengchen Sam Yong}
\affiliation{
    \position{Undergraduate Student}
    \department{Computer Science}
    \institution{Carnegie Mellon University}            
    \streetaddress{5000 Forbes Ave.}
    \city{Pittsburgh}
    \state{PA}
    \postcode{15213}
    \country{USA} 
}
\email{myong@andrew.cmu.edu}

\author{Jonathan Aldrich}
\orcid{0000-0003-0631-5591}             
\affiliation{
  \position{Professor}
  \department{Institute for Software Research}              
  \institution{Carnegie Mellon University}            
  \streetaddress{5000 Forbes Ave.}
  \city{Pittsburgh}
  \state{PA}
  \postcode{15213}
  \country{USA}                    
}
\email{jonathan.aldrich@cs.cmu.edu}          


\begin{abstract}
  Recently, user-centered methods have been proposed to improve the design of programming languages. In order to explore what benefits these methods might have for novice programming language designers, we taught a collection of user-centered programming language design methods to a group of eight students. We observed that natural programming and usability studies helped the students refine their language designs and identify opportunities for improvement, even in the short duration of a course project.
\end{abstract}

\maketitle

\section{Introduction}

PLIERS (Programming Language Iterative Evaluation and Refinement System) is a process for programming language design that integrates user-centered methods with traditional formal methods~\cite{Coblenz2020:PLIERS}. By including formal methods in the process, PLIERS is intended to facilitate creation of sound languages that are also as usable as possible. In designing PLIERS, Coblenz et al. evaluated the method by applying it themselves to two different programming languages. We were interested in assessing the PLIERS process by having people who did not develop it apply some of its techniques. What usability problems could language designers identify in their language prototypes? What challenges would they face in applying the methods? 

Evaluating design processes is challenging. One might like to conduct a randomized controlled trial in which a collection of designers are recruited to design languages. Some of the designers would be taught the method being evaluated, and the rest of the designers might receive some kind of training in traditional methods, such as case study or benchmarking methodology. Then, the designers would be given design tasks, and then resulting languages would be evaluated in quantitative user studies. 

Unfortunately, such a study is impractical; language design is very expensive, taking months or years. Furthermore, there are many relevant variables in such a study (such as individual designer experience, preferences, and skills), and controlling for them seems unlikely. Instead, we sought to use a lighter-weight approach to obtain some insights regarding what happens when those other than the designers of PLIERS use it. We leveraged our context: because the last author was teaching a programming language prototyping course, we conducted our study in the context of the course. Thus, our evaluation takes the form of a case study, in which we applied PLIERS in our course context.

By studying PLIERS in the context of a course we showed how user-centered methods can be useful to novice programming language designers. The students were able to identify shortcomings in their designs and identify new directions for language design. For example, one language was presented in functional style, yet even a programmer who preferred functional programming ended up using imperative constructs, suggesting that an imperative approach might be worth investigating. Another observation was that in \textit{natural programming} studies (in which  participants are typically asked to write programs without giving them any training at all), corpora can be used as a substitute for live participants: rather than asking participants to write bespoke code, one can analyze text that people have written already. 

\section{Related Work}
The problem of taking programmers into account when designing programming languages was discussed by Newell and Card in 1985: ``Millions for compilers but hardly a penny for understanding human programming language use''~\cite{Newell1985:Prospects}. Language designers have to make hundreds of different decisions when designing programming languages, and although many of these decisions likely impact programmers' abilities to achieve their goals when using the languages, designers lack a satisfactory way to leverage user data to inform their designs.

To address this problem, Stefik and Hanenberg have argued for thorough and careful randomized controlled trials of programming languages~\cite{stefik2017methodological}. Another approach has been to consider cognitive science, for example leveraging theories of natural language text understanding~\cite{Pennington1987:Stimulus}. Yet another tool is the \textit{cognitive dimensions of notation}, which  provide a vocabulary for discussing tradeoffs in notations~\cite{green1996usability}. PLIERS~\cite{Coblenz2020:PLIERS} represents another approach to providing design guidance, focusing on adapting methods from HCI research to the process of programming language design. Since the only users of PLIERS have been the designers, our focus in this work is on seeing whether others might be able to use the same techniques in their own programming language designs.

\section{Procedure}

The course in which we conducted our study was open to students who had already completed courses in either object-oriented programming or systems programming. The course centered around a project, in which individuals or small groups (up to two students per group) were asked to design a new programming language according to their own design goals. In addition, students were expected to complete assignments regarding programming language design and implementation.

Work on the project was divided into phases. After each phase, students received feedback from the course staff that they could leverage in future phases. Phases were as follows:

\begin{enumerate}
    \item \textit{Language proposal}, in which students proposed a new programming language for some domain. Students were told that they could choose a very narrow domain, such as ballet choreography, or a broad one, such as systems programming.
    \item \textit{Concepts for language design}, in which students applied Jackson's Design by Concept techniques~\cite{Jackson2019:Design} to produce an initial set of concepts for their language.
    \item \textit{Language semantics}, in which students defined abstract syntax and either static or dynamic semantics for fragments of their languages.
    \item \textit{User study design}, in which students proposed and prototyped a user study of key aspects of their programming languages.
    \item \textit{User study execution}, in which students ran a revised version of their user study.
    \item \textit{Final design and prototyping}, in which students revised their design based on the results of their study, implemented their language, and reported on the changes they had made based on the user studies.
\end{enumerate}

In addition to attending lectures on methods for completing the above work, students attended lectures and completed assignments regarding implementing interpreters for functional languages and transpilers for object-oriented languages. 

We taught students how to integrate user-centered methods into their language design process in four 80-minute class meetings. Although the authors presented slides, some of the time was spent discussing the material and letting the students try the techniques on small examples. We taught key concepts and methods that PLIERS suggests when designing programming languages:

\begin{enumerate}
    \item Defining usability properties of languages
    \item Recruitment techniques
    \item Participant selection and pre-screening
    \item Natural programming studies~\cite{naturalprogramming}
    \item Usability studies
        \begin{enumerate}
            \item Choosing appropriate usability questions
            \item Techniques for designing tasks in programming language studies, such as Wizard of Oz~\cite{dahlback1993wizard}
            \item Collecting data (e.g. think-aloud protocol)
            \item Post-study surveys
        \end{enumerate}
    \item Randomized controlled trials
\end{enumerate}

We received IRB approval to use the work products of the students for research purposes. Students obtained informed consent from their user study participants, whom they recruited as a convenience sample.

\section{Projects}

In order to put the results in context, we summarize in Table \ref{tab:projects} the projects that the students selected. To improve students' anonymity, we have replaced the language names with numeric identifiers (L1, L2, etc.). Although we taught a wide variety of techniques, we encouraged the students to consider either a usability study or a natural programming study in order to obtain results with an amount of effort that would be appropriate in the context of a course project.
\newcolumntype{L}[1]{>{\raggedright\let\newline\\\arraybackslash\hspace{0pt}}m{#1}}

\begin{table}[b]
    \centering
    \begin{tabularx}{\linewidth}{l X l >{\raggedright\arraybackslash}p{2cm}}
        \toprule
         \textbf{Name} & \textbf{Purpose} & \textbf{Students} & \textbf{Methods}\\
         \midrule
         L1 & A functional language for writing concurrent code & 2 & Natural programming\\ 
         L2 & Programming automatic knitting machines (for textile fabrication) & 1 & Natural programming \\ 
         L3 & Making it easier to write CSS code in BEM format~\cite{BEM} & 1 & Usability study \\ 
         L4 & Analogical reasoning, e.g. ``what is equivalent to apple if lemon is sour?'' & 1 & Natural programming \\ 
         L5 & Programming Internet of Things devices & 1 & Usability study \\ 
         L6 & Describing two-dimensional puzzles, such as Sudoku & 2 & Natural programming; usability study \\ 
         \bottomrule
    \end{tabularx}
    \caption{Language design projects selected by the students.}
    \label{tab:projects}
\end{table}

One additional student started a project on random number generation, but did not complete the project.

\section{Results and Discussion}

We organize this section by method, and summarize the experiences the students reported from the methods. For each method, we derive recommendations for future users of that method on programming language design.

\subsection{Natural Programming}

Projects L1, L2, L4, and L6 included natural programming studies. In L1, the students were interested in how programmers could specify operations on channels, which can be used in concurrent contexts to send messages between threads or processes. The students were inspired by Go~\cite{Donovan2015:Go}, which supports \code{<-} syntax for sending messages along channels. For example, in Go, the code below~\cite{go-channel-example} creates a channel called \code{messages}. Then, it sends the ``ping'' message along the channel \code{messages}. Finally, it receives a message along the \code{messages} channel, putting the result in \code{msg}.

\begin{lstlisting}[language=Go, xleftmargin=3ex]
messages := make(chan string)
go func() { messages <- "ping" }()
msg := <- messages
\end{lstlisting}

The students did not find support for such syntax in their natural programming study. Instead, participants invoked \code{send} and \code{receive} functions. Although this does not imply that the arrow syntax is not helpful, it at least suggests that using named functions may be more natural. This raises a question of the natural programming technique: for an approach to be \textit{natural}, is it necessary for participants to invent it themselves, or does it suffice for the technique to be \textit{easily learnable} with good results after learning? Perhaps \code{<-} is equally ``natural'' even though the participants were biased in their selection due to their past experience with other languages. Might there be substantive benefits of adopting syntactic choices that participants did not invent? The \code{<-} operator is indeed shorter, but the benefit may be so small that it cannot be practically measured, whereas \code{send} and \code{receive} might be more learnable. The tradeoff between concise and natural syntactic choices motivates further research to evaluate in what cases choices that are \textit{natural} are also better in practical ways, such as \textit{learnability}. Studies of non-native English speakers suggest that keyword choice may not be a very large factor in language usability~\cite{Reestman2019:Native}.

Project L2 conducted a variation on the natural programming technique. Due to limitations imposed by conducting the study during the COVID-19 pandemic, and the need to work with knitting domain experts, the student decided to conduct a corpus study rather than recruiting participants. Although corpus studies are normally not also natural programming studies, in this case, the corpus contained instructions that had been written for humans, not for machines. By mining Stitch-Maps~\cite{Stitch-Maps}, a web site hosting knitting patterns, it was possible to obtain a collection of five sets of instructions for knitting \textit{cables}, which are a hard-to-describe aspect of knitting instructions. Each set of instructions followed the same pattern, which was evidence for using the corresponding structure in the knitting programming language: \textit{Slip <loop count> to hold on <front/back>, stitch instructions on main row, stitch instructions on held stitches.} The corpus study was limited to patterns by one particular designer. Future work could expand the study to multiple designers to assess whether the same patterns are used by the broader knitting community.

The corpus also contained patterns that led to the following language constructs, which reference the domain-specific concepts of \textit{needles} and \textit{loops}:

\begin{sloppypar}
\begin{itemize}
\item Named needles or LIFO queues of loops are created by the phrase: \code{needle <identifier>}
\item A \textit{to} phrase, which places loops on an identified needle: \code{{operations} to <needle-identifier> in <front/back>}
\item A \textit{from} phrase, which consumes loops from an identified needle: \code{{operations} from
<needle-identifier>}
\end{itemize}
\end{sloppypar}
 
Project L4 conducted a natural programming study to develop a language for analogical reasoning. The student first asked participants to do some analogical reasoning themselves. For example, one prompt was:

\begin{quote}
    Bakers: the watchmakers of [NOUN]. You'll find them working on watches of their own: [VERB] the [NOUN], [VERB] the [NOUN], and [VERB] the [NOUN].
\end{quote}

Some participants gave \textit{bread} for the nouns, which is not surprising, but some gave interesting answers for the verbs, such as answers relating to \textit{winding}, which pertains to watchmaking. 

Then, participants were asked to propose a syntax for analogical queries. Although some proposed a form that was analogous to what they had been given, an interesting proposal to write a query for \textit{The integral, like a [NOUN], [VERB] space into tiny [NOUN:PLURAL]} was \code{integral, ? : ? -> ?}. That is, if an integral acts on \textit{something}, what else acts on something else in an analogous fashion?

The student reflected on the experience:

\begin{quote}
     I’m glad I did a natural programming study rather than a usability study of an existing system (which doesn't exist yet anyway), since it afforded a closer look into what kinds of analogies would be useful to query for\ldots and what expectations people have for such a system. Staying ``close to the metal'' of English and writing questions, rather than presenting a monolithic system outright, gave me more insight into the needs and quirks of potential users. It helped avoid the bias of a more constrained set of ``problems'' that I'd choose for a concrete usability study.
\end{quote}
 
L6 pertained to descriptions of two-dimensional puzzles. In one task, participants were asked to write code to describe a series of pictured grids. They used a variety of approaches: a C-like 2D array; a matrix constructor; and Python range syntax, including negative numbers for reverse indexing. The students observed that the participants seemed to be significantly influenced by their prior programming experience. In cases like these, other methods must be used to evaluate whether one approach is significantly better than the others, but these methods are likely not feasible in the short duration of the class project.
 


    

\subsection{Usability Studies}

Projects L3, L5, and L6 included usability studies. 

In L3, the student conducted an Internet-based usability study of a language to make it easier for CSS developers to use block-element-modifier conventions in their languages~\cite{BEM}. In the study, one task gave participants a pair of images, which represented two style variants of the same web page. Participants were asked to write code in the language to specify the styles. The primary usability problem the student identified was that participants with little general programming experience neglected to use the \code{@mod} feature (as was required) to express the variations in styles between the alternatives. However, it is not clear \textit{why} the participants made this mistake. This difficulty highlights a tradeoff in the design of usability studies. By conducting the study on the Internet, the student was able to attract a more diverse population of participants, but the format did not allow for think-aloud or an opportunity for the experimenter to ask questions of the participants. The latter may be a key ingredient in early-stage user studies in order to elucidate the causes of errors.

The student developing L5 asked participants to do three programming tasks. Participants were given example code and architectural documentation for the language, which facilitated programming distributed Internet-of-Things systems. The intent was to study how programmers could specify \textit{when} code would run, since code needed to run in response to various events. However, participants found the tasks more challenging than expected, in part because they found they needed to also address the question of \textit{where} the code would run in the distributed system. The study helped in identifying this problem, according to the student who designed the language:

 \begin{quote}
     Many questions were raised of the delineation of where code runs when writing a timing block. This was surprising as I believed the main issue \ldots was how to structure timing constructs\ldots.
\end{quote}

These observations helped motivate structural changes in the language. In the revised version, the questions of \textit{when} code runs is intertwined with the question of \textit{where} the code runs. Before, the language treated the two concepts as being orthogonal. The designers hope that the new approach will be easier to reason about.

The L6 study included a task in which participants were shown basic constructs of the language through an example, and then asked to use that language to describe Sudoku. The example showed functional-style code. Interestingly, one participant, who said their favorite language was SML (which is a functional language), wrote imperative-style code (including \code{loop} and \code{return}). It is notable  that the participant invented syntax seemingly at odds with their own preferences and with the example. Also, the programming task study changed smoothly into a natural programming study, in which the participant invented their own syntax. This suggests that in the future, it might be useful for study designers to think of programming tasks along a continuum from completely natural programming (in which participants receive no guidance at all) to a completely constrained environment, in which participants must satisfy a formal language specification.

\section{Conclusion and Future Work}

We derive several recommendations for language designers who seek to integrate user-centered design approaches into their work. For natural programming studies:

\begin{itemize}
    \item Using natural programming for selecting keywords is only a first step; naturalness is only one of several relevant criteria, and it is not yet known how to weigh the impact of naturalness.
    \item Consider using existing corpora as sources of data in natural programming studies. These have potentially lower cost and allow obtaining data from more diverse populations than one might otherwise have access to.
    \item Natural programming offers opportunities for deeper insights into the expectations and existing skills of user than might be obtained in a usability study. Usability studies focus participants on artifacts that may be unnatural for them; in contrast, natural programming studies consider participants as they are already.
    \item In cases where prior experience strongly impacts behavior, natural programming may be useful for identifying which experience is relevant, allowing development of tools that are more natural for particular groups.
\end{itemize}

For usability studies:
\begin{itemize}
    \item If the experimenter only has the results of participants' work, it can be impossible to infer why participants behaved they way they did. Think-aloud studies or asking follow-up questions of participants are likely better ways of understanding participant behavior.
    \item Usability studies can be useful for checking designers' assumptions about which parts of the programming tasks will be challenging. Parts that seem hard to the designer may not be significant obstacles to success, but other parts may represent serious usability problems.
    \item Even in an interactive user study, we need better techniques for understanding how people make decisions. In L6, it would be useful to know why a functional programmer chose an imperative approach, but we lack tools to answer that question definitively. Of course, if the experimenter had asked the participant, that would have been helpful, but experimenters frequently do not identify some interesting questions until it is too late to ask them. We can gain some insight from grounded theory methods, which leverage iteration in the analysis step; initial data from a study may help identify questions to ask participants in later iterations of the study.
\end{itemize}

This research focused on how beginning language designers might benefit from user-centered methods, particularly usability studies and natural programming studies. It identified several benefits of using these methods, but future work should investigate to what extent these methods are useful for experienced programming language designers. Future work should also investigate to what extent these methods lead to \textit{generalizable} language design guidelines, as opposed to results that are specific to the particular languages being studied. Finally, although experimenter skill and experience certainly plays a role, methodological development may improve the ability of experimenters to capture deeper insights about programmer behavior.

\begin{acks}
We are grateful for the assistance of the participants in our user studies.
\end{acks}

\bibliographystyle{ACM-Reference-Format}
\bibliography{obsidian}

\end{document}